\documentclass{aipproc}
\layoutstyle{8x11single}

\usepackage[T1]{fontenc}
\RequirePackage{times}
\RequirePackage{mathptm}
\usepackage{graphicx}
\begin{document}

\title{Characterization of quantum computable decision problems by state discrimination}

\author{Karl Svozil}{address={Institute of Theoretical Physics, Vienna
    University of Technology, Wiedner Hauptstra\ss e 8-10/136, A-1040
    Vienna, Austria},email={svozil@tuwien.ac.at}}



\begin{abstract}
One advantage of quantum algorithms over classical computation
is the possibility to spread out, process, analyse and extract information
in multipartite configurations in coherent superpositions of classical states.
This will be discussed
in terms of quantum state identification problems
based on a proper partitioning of mutually orthogonal sets of states.
The question arises whether or not it is possible
to encode equibalanced decision problems
into quantum systems, so that a single invocation
of a filter used for state discrimination suffices to obtain the result.
\end{abstract}


\maketitle

\section{Outline}

The question as to what might be considered the ``essence'' of quantum computation,
and its possible advantages over classical computation, has been the topic of
numerous considerations, both from a physical
(e.g., Ref.~\cite{ekerj96,pres-97,pres-ln,nielsen-book,galindo-02,mermin-04,eisert-wolf-04})
as well as from a computer science
(e.g., Ref.~\cite{Gruska,benn:97,Ozhigov:1997,bbcmw-01,cleve-99,fortnov-03}) perspective.
Contributing to this ongoing research,
we will present an analysis of novel propositional structures
in quantum mechanics; i.e.,
on the issue of what kind of propositions about
quantum computers exist which do not correspond to any classical statement.
We will consider coherent superpositions of  states and will
make explicit use of the fact that in quantum mechanics information
can be coded in or ``spread among'' entangled multipartite systems in such a way that
information about the single quanta is not useful for (and even makes impossible)
a decryption of the quantum computation.

Alas, it is quite evident that not all decision problems
have a proper encoding
into some quantum mechanical system
such that their resources (computation time, memory usage) is bound by some
criterion such as polynomiality or even finiteness.
Take, as a concrete example, a particular type of halting problem:
Alice presents Bob a black box with input and output interfaces.
Bob's task is to find out whether an arbitrary function of $n$ bits encoded in the black box
will ever output "0."
As this configuration could essentially get as worse as a {\em busy beaver} problem
\cite{rado},
the time it takes for Alice's box to ever output a "0" may grow faster than
any recursive (i.e., computable \cite{rogers1,odi:89}) function of $n$.

Is it possible to
characterize the exact domain of functions
and propositions about them which can be ``reasonably''
(e.g., polynomially) coded into a quantum computation,
given an fairly general set of coding strategies, such as unitary transformations?
In what follows, an attempt is made to characterize the class of quantum computable functions whose
computational complexity grows {\em linearly} with the number of bits
by considering partitioning of states and the associated propositions and observables
\cite{zeil-99,DonSvo01,svozil-2002-statepart-prl,svozil-2003-garda}.
Certain quantum computations such as the Deutsch algorithm
will be expressed as state identification problems,
resulting in the systematic construction of a great variety of computations
corresponding to (incomplete) state identifications
based on superposition and interference.

The notation of Mermin~\cite{mermin-02,mermin-04,mermin-qc} will be adopted.
Consider at first a single qubit in its most general form
$
\vert \psi \rangle
=
\alpha_0
\vert 0 \rangle
+
\alpha_1
\vert 1 \rangle
$
with
$
\vert
\alpha_0
\vert^2
+
\vert
\alpha_1
\vert^2
=1$
as a coherent superposition
between some ``quasi-classical''
states
$
\vert 0 \rangle
$ and
$
\vert 1 \rangle
$
of the computational basis representable by the set of orthogonal vectors
$\{\vert 0 \rangle \equiv (1,0)^T,\vert 1 \rangle \equiv (0,1)^T\}$
(the superscript $T$ indicates transposition).
A 50:50 mixture of the quasi-classical states
is obtained by
$\textsf{\textbf{H}} \vert 0 \rangle
=
(1/\sqrt{2})\left(
\vert 0 \rangle
+
\vert 1 \rangle
\right)
$ or $\textsf{\textbf{H}} \vert 1 \rangle
=
(1/\sqrt{2})\left(
\vert 0 \rangle
-
\vert 1 \rangle
\right)
$
where $\textsf{\textbf{H}}$ is the normalized Hadamard matrix
$\frac{1}{\sqrt{2}}\left(
\begin{array}{rr}
1&1\\
1&-1\\
\end{array}
\right)$.
According to quantum logic
\cite{birkhoff-36,v-neumann-55,svozil-ql},
the interpretation of
$
\textsf{\textbf{H}} \vert 0 \rangle
$
or
$
\textsf{\textbf{H}} \vert 1 \rangle
$
it is the proposition,
{\em ``the quant is in the state associated with the projector}
$(1/2)\left(\textsf{\textbf{1}} \pm \textsf{\textbf{X}}\right)$,''
where $\textsf{\textbf{1}}$ is the unitity and $\textsf{\textbf{X}}
=
\left(
\begin{array}{rr}
0&1\\
1&0\\
\end{array}
\right)$ is the {\em not}-operator.
Classically, neither these states nor the projectors correspond to any
opertionalizable physical entity.
Quantum mechanically, they have,
for instance, an interpretation in terms of electron or neutron spin states
and spin state measurements by a Stern-Gerlach apparatus,
or in terms of photon polarization states and polarization measurements.
Since
$(1/2)(\textsf{\textbf{1}} \pm  \textsf{\textbf{X}})=
(1/2)\left[\textsf{\textbf{1}} + {\bf \sigma}( \theta =\pm \pi /2,\varphi=0)\right]$
with
$
{\bf \sigma}( \theta ,\varphi )=
\left(
\begin{array}{rr} \cos \theta  &e^{-i\varphi} \sin \theta   \\
  e^{i\varphi}\sin \theta  & -\cos \theta
  \end{array}
\right)
$
for the polar angle $\theta$ and the azimuthal angle $\varphi$,
the physical proposition  corresponding to
$\textsf{\textbf{H}}\vert 0\rangle$ and
$\textsf{\textbf{H}}\vert 1\rangle$ is
{\em ``along the polar angle $\pm \pi/2$ and azimuthal angle $\varphi=0$,
the particle is in a linear polarization (or positive spin) state.''}

\section{Identifying states among contexts}

A {\em context} can formally be defined \cite{svozil-2004-vax}
as a single (nondegenerate) ``maximal'' self-adjoint operator
$\textsf{\textbf{C}}$.
It has a spectral
decomposition into some complete set of orthogonal projectors $\textsf{\textbf{E}}_i$
which correspond to propositions in
the von Neumann-Birkhoff type sense \cite{birkhoff-36,v-neumann-49}.
That is, $\textsf{\textbf{C}}=\sum_{i=1}^d e_i \textsf{\textbf{E}}_i$
with mutually different real $e_i$ and some orthgonal projectors
$\{\textsf{\textbf{E}}_i{} \mid i=1,\ldots d\}$ of
$d$-dimensional Hilbert space.
In $d$ dimensions, contexts can be viewed as $d$-pods or orthogonal bases
spanned by the vectors associated with the $d$ mutually orthogonal projectors
$\textsf{\textbf{E}}_1,
\textsf{\textbf{E}}_2, \cdots, \textsf{\textbf{E}}_d$.

The general problem to (uniquely) identify orthogonal pure states
among contexts resulting from $k$ particles in $n=2$ or more dimensions per particle
has been solved in Ref.~\cite{DonSvo01,svozil-2002-statepart-prl,svozil-2003-garda}
{\em via} a system of $k$ co-measurable  filters $\textsf{\textbf{F}}_i$, $i=1,\ldots, k$
with the following properties:
\begin{enumerate}
\item[(F1)]
Every filter $\textsf{\textbf{F}}_i$
corresponds to an operator (or a set of operators)
which generates an
equi-$n$-partition of the $d$-dimensional state space into
$n$ slices (i.e., partition elements) containing $d/n=d^{1-1/k}=n^{k-1}$ states per slice.
(Note that $d=n^k$.)
A filter is said to separate two eigenstates if the eigenvalues are different.
\item[(F2)]
From each one of the $k$ partitions of (F1), take an arbitrary element.
The intersection of the elements of all different partitions
results in a {\it single} one of the $d=n^k$ different states.
\item[(F3)]
The union of all those single states generated by the intersections of (F2)
is the entire set of states.
\end{enumerate}

For $n=2$, an explicit construction of all the systems of filters and their associated propositions
can be given in terms of projectors and their orthogonal projectors;
every one of them projecting onto a $d/2$-dimensional subspace,
such that the serial composition of any complete set of (orthogonal) projectors (one per filter)
yields the finest resolution; i.e., some of the $d$ one-dimensional projectors $\textsf{\textbf{E}}_i$
spanning the context $\textsf{\textbf{C}}$.

The system of filters resolving $\textsf{\textbf{C}}$
is not unique;
all such systems of filters can be obtained by permutating
the columns of the matrix whose rows are
the diagonal elements of all the filters in diagonalized form.
Different contexts
$\textsf{\textbf{C}}'$
are resolved by different systems of filters which are obtained by transforming
$\textsf{\textbf{F}}_i$, $i=1,\ldots, k$ through
the same basis transformation which transforms
$\textsf{\textbf{C}}$
into
$\textsf{\textbf{C}}'$.
Several examples and explicit constructions will be given below.

Take, for instance, three two-state quanta, i.e.,
the case $k=3$, $n=2$, and thus $d=2^3$.
The three projectors
$$
\begin{array}{ccc}
\textsf{\textbf{F}}_1&=&\textrm{diag}(1,1,1,1,0,0,0,0),\\
\textsf{\textbf{F}}_2&=&\textrm{diag}(1,1,0,0,1,1,0,0),\\
\textsf{\textbf{F}}_3&=&\textrm{diag}(1,0,1,0,1,0,1,0),\\
\end{array}
$$
 together with their orthogonal projectors
$$
\begin{array}{ccc}
\textsf{\textbf{F}}_1'&=&\textrm{diag}(0,0,0,0,1,1,1,1), \\
\textsf{\textbf{F}}_2'&=&\textrm{diag}(0,0,1,1,0,0,1,1), \\
\textsf{\textbf{F}}_3'&=&\textrm{diag}(0,1,0,1,0,1,0,1),   \\
\end{array}
$$
form the system of three filters
$
\{
\{ \textsf{\textbf{F}}_1,\textsf{\textbf{F}}_1' \},
\{ \textsf{\textbf{F}}_2,\textsf{\textbf{F}}_2' \},
\{ \textsf{\textbf{F}}_3,\textsf{\textbf{F}}_3' \}
\}
$
which have the desired properties (F1)--(F3).
Equivalent filters are obtained by permuting the columns of the diagonal rows of
\begin{equation}
\left(
\begin{array}{cccccccc}
1&1&1&1&0&0&0&0\\
0&0&0&0&1&1&1&1\\
\hline
1&1&0&0&1&1&0&0\\
0&0&1&1&0&0&1&1\\
\hline
1&0&1&0&1&0&1&0\\
0&1&0&1&0&1&0&1\\
\end{array}
\right).
\label{2005-ko-e78}
\end{equation}
Different systems of filters are obtained by permutating the columns
of the matrix in Eq.~\ref{2005-ko-e78}; e.g.,
\begin{equation}
\left(
\begin{array}{cccccccc}
1&1&1&0&0&0&0&1\\
0&0&0&1&1&1&1&0\\
\hline
1&0&0&1&1&0&0&1\\
0&1&1&0&0&1&1&0\\
\hline
0&1&0&1&0&1&0&1\\
1&0&1&0&1&0&1&0\\
\end{array}
\right),
\qquad
\left(
\begin{array}{cccccccc}
1&1&0&0&0&0&1&1\\
0&0&1&1&1&1&0&0\\
\hline
0&0&1&1&0&0&1&1\\
1&1&0&0&1&1&0&0\\
\hline
1&0&1&0&1&0&1&0\\
0&1&0&1&0&1&0&1\\
\end{array}
\right),
\qquad
\ldots
\label{2005-ko-e781}
\end{equation}
In the case of $k=2$, any permutation yields the original system of filters.

Different contexts are reached by transforming every single filter operator
through the same unitary transformation.
Note that the row permutations and unitary transformations are exhaustive;
i.e., there are no other methods available.
For $n>2$, the filter operators cannot correspond to projectors, because
they are not binary but $n$-ary.
In this case,  for instance, $n^k$ different
prime numbers can be used as eigenvalues.
A more detailed treatment of this case can be found
in Refs.~\cite{svozil-2002-statepart-prl,svozil-2003-garda}.

\section{Deutsch's problem and related algorithms}

In what follows, Deutsch's decision problem to find out whether or not an unknown
function $f$ that takes a single (classical) bit into a single (classical) bit
is constant or not, which is equal to finding the parity of $f:\{0,1\}\rightarrow \{0,1\}$, will
be interpreted as a state identification problem, which is solved by the methods
discussed in the previous section.
There are four possible bivalent functions of one bit:
the constant functions $f_0$ and $f_3$ take
any bit value and map it into either $0$ or $1$, respectively.
The two remaining functions $f_1$ and $f_2$
correspond to the identity $\textsf{\textbf{1}}$
and to the {\em not} operator $\textsf{\textbf{X}}$, and are thus not constant
(cf.~Table~\ref{2005-ko-t11dp}).
\begin{table}
\centerline{
\begin{tabular}{cccccccccccccc}
\hline
$f$& $ 0$ &$1$\\
\hline
$f_0$ &$0$ &$0$\\
$f_1$ &$0$ &$1$\\
$f_2$ &$1$ &$0$\\
$f_{3}$ &$1$ & $1$\\
\hline
\end{tabular}
}
\caption{The binary functions of one bit considered in Deutsch's problem. \label{2005-ko-t11dp}
}
\end{table}
Hence, with respect to constancy, the set of all functions
$\{f_0,f_1,f_2,f_3\}$
is equipartitioned into
\begin{equation}
F_D=\{\{f_0,f_3\},\{f_1,f_2\}\}.
\label{2005-ko-03}
\end{equation}
The first and second elements   $\{f_0,f_3\}$ and $\{f_1,f_2\}$ of this partition
can be interpreted as the proposition,
{\em ``the function is (not) constant.''}

When coding the Deutsch problem and the computation of $f$
into a state identification problem,
one task is to map the binary partition $F_D$ in Eq.~(\ref{2005-ko-03})
into a quantum state filter
$\textsf{\textbf{F}}$ with equivalent separation properties.
Presently, there does not exist any algorithmic way (only heuristic ones)
to obtain such a quantum encoding,
nore is any one likely to exist
(cf. the parity problem discussed below).

First note that, as the functions $f_0$ and $f_3$ are two-to-one (i.e., irreversible),
the input bit needs to be augmented by a second bit to maintain
reversibility, which is a necessary condition for the  unitarity of the state evolution.
Usually, this is accomplished by considering
$\textsf{\textbf{U}}_f(\vert x \rangle \vert y \rangle )=\vert x \rangle \vert y \oplus f(x)\rangle$, where $\oplus$ is the modulo-2 addition
(without carrying).

The encoding Ansatz enumerated in Table~\ref{2005-ko-t1} represents the evolution
of the single terms contributing to
$\textsf{\textbf{U}}_f(\textsf{\textbf{H}}\otimes\textsf{\textbf{H}})
(\textsf{\textbf{X}}\otimes\textsf{\textbf{X}})
(\vert 0\rangle \vert 0\rangle )$,
resulting in the two different states
\begin{equation}
\vert \psi_1 \rangle = \pm \frac{1}{2} (
\vert 0\rangle   -
\vert 1\rangle  )(
\vert 0\rangle   -
\vert 1\rangle  )\equiv \pm \frac{1}{2} ((1,-1)\otimes (1,-1))^T=\pm \frac{1}{2} (1,-1,-1,1)^T
\label{2005-ko-e111}
\end{equation}
for $f_0$ as well as $f_3$, and
\begin{equation}
\vert \psi_2 \rangle = \pm \frac{1}{2} (
\vert 0\rangle   +
\vert 1\rangle  )(
\vert 0\rangle   -
\vert 1\rangle  )\equiv \pm \frac{1}{2} ((1,1)\otimes (1,-1))^T=\pm \frac{1}{2} (1,-1,1,-1)^T
\label{2005-ko-e112}
\end{equation}
for $f_1$ as well as $f_2$.
\begin{table}
\centerline{
\begin{tabular}{cccccccccccccc}
\hline
 & $\frac{1}{2}\big[\vert 0\rangle \vert 0 \oplus f(0)\rangle $ &$-$& $\vert 0\rangle \vert 1 \oplus f(0)\rangle $ &$-$& $\vert 1\rangle \vert 0 \oplus f(1)\rangle $ &+& $\vert 1\rangle \vert 1 \oplus f(1)\rangle \big]$\\
\hline
$f_0$:  & $\frac{1}{2}\big(\vert 0\rangle \vert 0\rangle $ &$-$& $\vert 0\rangle \vert 1\rangle $ &$-$& $\vert 1\rangle \vert 0\rangle $ &+& $\vert 1\rangle \vert 1\rangle \big)$\\
$f_1$:  & $\frac{1}{2}\big(\vert 0\rangle \vert 0\rangle $ &$-$& $\vert 0\rangle \vert 1\rangle $ &$-$& $\vert 1\rangle \vert 1\rangle $ &+& $\vert 1\rangle \vert 0\rangle \big)$\\
$f_2$:  & $\frac{1}{2}\big(\vert 0\rangle \vert 1\rangle $ &$-$& $\vert 0\rangle \vert 0\rangle $ &$-$& $\vert 1\rangle \vert 0\rangle $ &+& $\vert 1\rangle \vert 1\rangle \big)$\\
$f_3$:  & $\frac{1}{2}\big(\vert 0\rangle \vert 1\rangle $ &$-$& $\vert 0\rangle \vert 0\rangle $ &$-$& $\vert 1\rangle \vert 1\rangle $ &+& $\vert 1\rangle \vert 0\rangle \big)$\\
\hline
\end{tabular}
}
\caption{State evolution of $\textsf{\textbf{U}}_f(\textsf{\textbf{H}}\otimes\textsf{\textbf{H}})
(\textsf{\textbf{X}}\otimes\textsf{\textbf{X}})
(\vert 0\rangle \vert 0\rangle )$
for the four functions $f_0,f_1,f_2,f_3$. \label{2005-ko-t1}
}
\end{table}
Together with
$\vert \psi_3 \rangle = (\textsf{\textbf{H}}\otimes\textsf{\textbf{H}})(\vert 0\rangle \vert 0\rangle )\equiv
(1/2)(1,1,1,1)^T$
and
$\vert \psi_4 \rangle = (\textsf{\textbf{H}}\otimes\textsf{\textbf{H}})(\textsf{\textbf{X}}\otimes\textsf{\textbf{1}})(\vert 0\rangle \vert 0\rangle )\equiv
(1/2)(1,1,-1,-1)^T$,
the four states in $\textbf{B}^D=\{\psi_1, \psi_2 ,\psi_3 ,\psi_4 \}$ form an orthonormal basis.

Application of two Hadamard-transformations for each one of the two bits finally yields a
representation in the sandard computational basis; i.e.,
\begin{equation}
(\textsf{\textbf{H}}\otimes\textsf{\textbf{H}})
\textsf{\textbf{U}}_f
(\textsf{\textbf{H}}\otimes\textsf{\textbf{H}})
(\textsf{\textbf{X}}\otimes\textsf{\textbf{X}})
(\vert 0\rangle \vert 0\rangle )
=
\left\{
\begin{array}{ccl}
\vert 1\rangle \vert 1\rangle \equiv (0,0,0,1)^T  &\textrm{ for } f(0)=f(1),\\
\vert 0\rangle \vert 1\rangle \equiv (0,1,0,0)^T   &\textrm{ for } f(0)\neq f(1).\\
\end{array}
\right.
\label{2005-ko-e113}
\end{equation}

We are now in the position to formulate the state identification problem corresponding to the Deutsch
algorithm.
This is achieved
by considering the projector
$\textsf{\textbf{F}}_1 =\textrm{diag}(1,1,0,0)$,
which, together with its orthogonal projector
$\textsf{\textbf{F}}_1' =\textrm{diag}(0,0,1,1)$,
constitutes a filter
corresponding to the binary partition $F_D$ in Eq.~(\ref{2005-ko-03}).
Note that a second filter $\textsf{\textbf{F}}_2$, based on the projections
$\textsf{\textbf{F}}_2 =\textrm{diag}(1,0,1,0)$
and
$\textsf{\textbf{F}}_2' =\textrm{diag}(0,1,0,1)$, completes the system of filters.
It is unable to separate
$\vert 1 1\rangle$
from
$\vert 0 1 \rangle$, but separates
$\vert 0 0 \rangle$
and
$\vert 1 0 \rangle$
from
$\vert 0 1\rangle$
and
$\vert 1 1\rangle$.

Alternatively, we may consider the state identification problem without the final Hadamard transformations
as, {\em ``find the observables which separate $\psi_1$ from $\psi_2$.''}
The complete state identification problem should also contain the observables separating $\psi_3$ from $\psi_4$,
but in Deutsch's problem one is not primarily interested
in uniquely identifying the function itself; rather in its (non)constancy.
Hence, it is not necessary to employ the entire system of two filters, but rather
a single filter constructed to separate $f_0$, $f_3$ from  $f_1$, $f_2$.
This is achieved
by transforming the two operators
$\textsf{\textbf{F}}_1 =\textrm{diag}(1,1,0,0)$
and
$\textsf{\textbf{F}}_2 =\textrm{diag}(1,0,1,0)$
associated with a binary search type state separation in the  basis
$\textbf{B}=\{
(1,0,0,0)^T,
(0,1,0,0)^T,
(0,0,1,0)^T,
(0,0,0,1)^T\}$
through
$\textsf{\textbf{U}}\textsf{\textbf{F}}_1\textsf{\textbf{U}}^{-1} =\textsf{\textbf{F}}^D_1$ and
$\textsf{\textbf{U}}\textsf{\textbf{F}}_2\textsf{\textbf{U}}^{-1} =\textsf{\textbf{F}}^D_2$,
where
\begin{equation}
\textsf{\textbf{U}}=\frac{1}{2}
\left(
\begin{array}{rrrr}
 1&  1& 1&  1\\
 1& -1& 1& -1\\
-1&  1& 1& -1\\
-1& -1& 1&  1
\end{array}
\right)
\end{equation}
is the unitary transformation which corresponds to a basis change
$\textbf{B} \rightarrow \textsf{\textbf{U}}\textbf{B}=\textbf{B}^D$.
It is straightforward to check that, by the eigenvalue spectrum,
$\textsf{\textbf{F}}^D_1$ separates between
$\psi_1$ and $\psi_3$ from
$\psi_2$ and $\psi_4$
(and at the same time,
$\textsf{\textbf{F}}^D_2$ separates between
$\psi_1$ and $\psi_2$ from
$\psi_3$ and $\psi_4$).
Hence, $\textsf{\textbf{F}}^D_1$ generates a partition
$\{\{\psi_1,\psi_3\},\{\psi_2,\psi_4\}\}$ of the set
$\{\psi_1,\psi_3,\psi_2,\psi_4\}$
of orthogonal states.
($\textsf{\textbf{F}}^D_2$ generates the partition
$\{\{\psi_1,\psi_2\},\{\psi_3,\psi_4\}\}$.)
The states $\psi_i$, however, do not directly correspond to the functions $f_j$ in the
Deutsch partition in Eq.~(\ref{2005-ko-03}); they rather represent
joint properties of these functions, such as constancy.

Another encoding strategy of the Deutsch problem
can be based upon a immediate identification of
$\{f_0,f_1,f_2,f_3\}$
with the four states of the computational basis $\textbf{B}$.
The nontrivial part in this case is the mapping of the functions $f_i$ on to $\textbf{B}$;
e.g., by constructing unitary transformations depending
on $f_i$ and acting on $\vert 0 0 \rangle$, such
as for instance $\textsf{\textbf{V}}_f\vert 0 0 \rangle=\vert f(0) f(1) \rangle$.
Once this has been achieved,
in order to express constancy, the filter would then have to
separate the orthogonal (Bell) states
$\varphi_{1,4}\equiv (1,0,0,\pm 1)^T$
from
$\varphi_{2,3}\equiv  (0,1,\pm 1,0)^T$;
a rather straightforward task.

Still another encoding strategy would be to invoke the phase oracle
$
\textsf{\textbf{U}}_f \left( \vert x\rangle  \otimes \textsf{\textbf{H}}
\vert 1\rangle \right)
=
(-1)^{f(x)}    \vert x\rangle   \otimes \textsf{\textbf{H}}
\vert 1\rangle $. The resulting states are enumerated in Table~\ref{2005-ko-t11}.
\begin{table}
\centerline{
\begin{tabular}{cccccccccccccc}
\hline
& \multicolumn{4}{c }{$(-1)^{f(x)}$}\\
$f$& $\vert 0\rangle$ &$\vert 1\rangle $\\
\hline
$f_0$ &$+$ &$+$\\
$f_1$ &$+$ &$-$\\
$f_2$ & $-$&$+$\\
$f_3$ &$-$ & $-$\\
\hline
\end{tabular}
}
\caption{The phase factors of $(-1)^{f(x)}\vert xy\rangle$. \label{2005-ko-t11}
}
\end{table}
The phases result in the orthogonality of the two linear
subspaces corresponding to $f_0$ and $f_3$, with respect to $f_1$ and $f_2$.

In a very similar manner, one could discuss the Bernstein-Vazirani algorithm,
 as well as
the Deutsch-Josza and  Simon's decision problems
(in the latter cases with the proviso discussed later,
since the algorithm is not deterministic).
Note that this method exhausts all possible decision problems based on equipartitioning
of state spaces, but does not give a direct hint
about the type of classical algorithmic problem which are solvable that way.

\section{Parity checking}

Deutsch's problem is just the simplest in a particular class of problems:
check the parity of an unknown binary function $f:\{0,1\}^k \rightarrow \{0,1\}$ of $k$ bits.
There are $2^{2^k}$ such functions.
The parity of a function $f$ of $k$ bits depends on whether
the number of functional values of $f(x_1,\ldots , x_k)=1$
on all $x_1,\ldots , x_k\in \{0,1\}$ is even or odd,
denoted by  ``$+$'' and ``$-$,'' respectively.

Consider, for the sake of an example,
two bits $x,y$ and an unknown function $f(x,y)$
of all the $2^{2^2}=16$ binary functions partly listed in Tab.~\ref{2005-ko-t2}.
\begin{table}
\centerline{
\begin{tabular}{cccccccccccccc}
\hline
 $\pm$&$f$&  00& 01& 10 &11\\
\hline
$+$&$f_0$ &0 &0& 0& 0\\
$-$&$f_1$ &0 &0& 0& 1\\
$-$&$f_2$ &0 &0& 1& 0\\
\multicolumn{6}{c}{$\cdots$}\\
$+$&$f_{15}$ &1 & 1& 1&1\\
\hline
\end{tabular}
}
\caption{Listing of the 16 binary functions of two variables $x,y$ with their parity bits
``$\pm$''. \label{2005-ko-t2}
}
\end{table}
The set of 16 functions can be equipartitioned
into two groups of 8 functions,  according to positive and negative parity; i.e.,
\begin{equation}
F_P=
\left\{
\left\{
     f_{  0} ,
     f_{  5} ,
     f_{  6} ,
     f_{  7} ,
     f_{  8} ,
     f_{  9} ,
     f_{ 10} ,
     f_{ 15}
\right\},
\left\{ f_{  1} ,
     f_{  2} ,
     f_{  3} ,
     f_{  4} ,
     f_{ 11} ,
     f_{ 12} ,
     f_{ 13} ,
     f_{ 14}
\right\}
\right\}
.
\label{2005-ko-e55-0}
\end{equation}
One might be tempted to speculate
that the corresponding proposition corresponds to some realizable
quantum filter which separates the two parity classes by some quantum implementation
$
\textsf{\textbf{U}}_f
$ in a single run.
Motivation for this comes from the direct and ``local,'' or ``isolated'' evaluation
of the functional values;
without any recursion, iteration, or additional functional and contextual relation
between the values.
Despite these indications, the parity of a function
has been proven quantum computationally hard
\cite{Farhi-98,bbcmw-01,Miao-2001,orus-04,stadelhofer-05}:
It is only possible to go from $2^k$ classical queries down to $2^k/2$
quantum queries, thereby gaining a factor of 2.

Classically, parity checking grows exponentially $2^k$ with the number $k$ of bits of the
functional arguments, as there is no other was than to compute the functional
values on the entire set of $2^k$ arguments.
Quantum mechanically,
one may interpret this problem as a particular instance
of a generalized Grover algorithm with an unknown number of special states,
which can be solved by applying the quantum Fourier transform.

By making use of the phase oracle
$
\textsf{\textbf{U}}_f \left( \vert x\rangle  \otimes \textsf{\textbf{H}}
\vert 1\rangle \right)
=
(-1)^{f(x)}    \vert x\rangle   \otimes \textsf{\textbf{H}}
\vert 1\rangle $,
one obtains, after a second application of a Hadamard transformation,
\begin{equation}
\left(
\textsf{\textbf{1}}
\otimes
\textsf{\textbf{1}}
\otimes
\textsf{\textbf{H}}
\right)
\textsf{\textbf{U}}_f
\left(
\textsf{\textbf{1}}
\otimes
\textsf{\textbf{1}}
\otimes
\textsf{\textbf{H}}
\right)
\vert x,y \rangle \vert 1\rangle
=
(-1)^{f(x,y)}\vert x,y \rangle \vert 1\rangle
.
\label{2005-ko-e55}
\end{equation}
Table \ref{2005-ko-t3} lists the results of this transformation.
\begin{table}
\centerline{
\begin{tabular}{ccccccccccccccc}
\hline
&& &\multicolumn{4}{c}{$(-1)^{f(x)} $}
\\
$\pm$&$f$& &
$\vert 00 \rangle$&
$\vert 01 \rangle$&
$\vert 10 \rangle$&
$\vert 11 \rangle$
\\
\hline
$+$&$f_{  0}      $ &        &$+$ &$+$ &$+$ &$+$                               \\
$-$&$f_{  1}      $ &        &$+$ &$+$ &$+$ &$-$                               \\
$-$&$f_{  2}      $ &        &$+$ &$+$ &$-$ &$+$                               \\
$-$&$f_{  3}      $ &        &$+$ &$-$ &$+$ &$+$                               \\
$-$&$f_{  4}      $ &        &$-$ &$+$ &$+$ &$+$                               \\
$+$&$f_{  5}      $ &        &$+$ &$+$ &$-$ &$-$                               \\
$+$&$f_{  6}      $ &        &$+$ &$-$ &$+$ &$-$                               \\
$+$&$f_{  7}      $ &        &$-$ &$+$ &$+$ &$-$                               \\
$+$&$f_{  8}      $ &        &$+$ &$-$ &$-$ &$+$                               \\
$+$&$f_{  9}      $ &        &$-$ &$+$ &$-$ &$+$                               \\
$+$&$f_{ 10}      $ &        &$-$ &$-$ &$+$ &$+$                               \\
$-$&$f_{ 11}      $ &        &$+$ &$-$ &$-$ &$-$                               \\
$-$&$f_{ 12}      $ &        &$-$ &$+$ &$-$ &$-$                               \\
$-$&$f_{ 13}      $ &        &$-$ &$-$ &$+$ &$-$                               \\
$-$&$f_{ 14}      $ &        &$-$ &$-$ &$-$ &$+$                               \\
$+$&$f_{ 15}      $ &        &$-$ &$-$ &$-$ &$-$                               \\
\hline
\end{tabular}
}
\caption{The phases from Eq.~(\ref{2005-ko-e55}).
\label{2005-ko-t3} }
\end{table}
As long as
the function is ``unbalanced,'' such that the
number of values of $f(x_1, \ldots ,x_k)=1$ is small compared to $2^k$,
a quadratic speedup is achievable.
However, this condition does in general not apply.

\section{Generalized Deutsch algorithms}

In what follows  we shall present a type of quantum algorithm
which is directly motivated by the state identification problem.
Consider the class of binary functions of two variables which are  the sums of
two (or more) binary functions of one variable; e.g.,
\begin{equation}
f_{ij}(x,y)=f_i(x)+f_j(y); \quad 0\le i,j\le 3.
\label{2005-ko-e31}
\end{equation}
The binary functions $f_i,f_j$ of one bit are the same as in Deutsch's problem
listed in Table~\ref{2005-ko-t11dp}.
The corresponding unitary transformations given by
$
\textsf{\textbf{U}}_{f_{ij}}=
\textsf{\textbf{U}}_{f_i}
\otimes
\textsf{\textbf{U}}_{f_j}$.
In this case,
the phase oracle  yields phases which are listed in Table~\ref{2005-ko-t31}.
\begin{table}
\centerline{
\begin{tabular}{ccccccccccccccc}
\hline
& &\multicolumn{4}{c}{$(-1)^{f_i(x)+f_j(y)} $}
\\
$f$& &
$\vert 00 \rangle$&
$\vert 01 \rangle$&
$\vert 10 \rangle$&
$\vert 11 \rangle$
\\
\hline
    $f_{  00}      $ &   &$+$&$+$&$+$&$+$                                     \\
    $f_{  01}      $ &   &$+$&$-$&$+$&$-$                                     \\
    $f_{  02}      $ &   &$-$&$+$&$-$&$+$                                     \\
    $f_{  03}      $ &   &$-$&$-$&$-$&$-$                                     \\
    $f_{  10}      $ &   &$+$&$+$&$-$&$-$                                     \\
    $f_{  11}      $ &   &$+$&$-$&$-$&$+$                                     \\
    $f_{  12}      $ &   &$-$&$+$&$+$&$-$                                     \\
    $f_{  13}      $ &   &$-$&$-$&$+$&$+$                                     \\
    $f_{  20}      $ &   &$-$&$-$&$+$&$+$                                     \\
    $f_{  21}      $ &   &$-$&$+$&$+$&$-$                                     \\
    $f_{  22}      $ &   &$+$&$-$&$-$&$+$                                     \\
    $f_{  23}      $ &   &$+$&$+$&$-$&$-$                                     \\
    $f_{  30}      $ &   &$-$&$-$&$-$&$-$                                     \\
    $f_{  31}      $ &   &$-$&$+$&$-$&$+$                                     \\
    $f_{  32}      $ &   &$+$&$-$&$+$&$-$                                     \\
    $f_{  33}      $ &   &$+$&$+$&$+$&$+$                                     \\
\hline
\end{tabular}
}
\caption{The phases from the phase oracle applied to Eq.~(\ref{2005-ko-e31}).
\label{2005-ko-t31} }
\end{table}

The four orthogonal vectors resulting from the phase enumeration in Table~\ref{2005-ko-t31}
form a basis
$
\textsf{\textbf{B}}'=\{
\varphi_1,
\varphi_2,
\varphi_3,
\varphi_4\}$, with
\begin{equation}
\begin{array}{lll}
\varphi_1&=&  (1,1,1,1)^T,\\
\varphi_2&=&  (1,1,-1,-1)^T,\\
\varphi_3&=&   (1,-1,1,-1)^T,\\
\varphi_4&=&   (1,-1,-1,1)^T.\\
\end{array}.
\label{2005-ko-e55-023}
\end{equation}

Consider the decision problems corresponding to the following propositions:
\begin{enumerate}
\item[(D1)]
{\em The function $f_{ij}(x,y)$  is  constant in the first argument.}

\item[(D2)]
{\em The function $f_{ij}(x,y)$  is  constant in the second argument.}

\item[(D3)]
{\em The function $f_{ij}(x,y)$  is  constant in the first argument and not constant in the second argument,
                           or it is  constant in the second argument and not constant in the first argument.}

\item[(D4)]
{\em The function $f_{ij}(x,y)$  is  constant in the first argument and constant in the second argument,
                           or it is  not constant in the second argument and not constant in the first argument.}
\end{enumerate}

The partitions corresponding to these decision problems are
\begin{eqnarray}
F_1&=&
\left\{
\left\{
     f_{ 00} ,
     f_{ 01} ,
     f_{ 02} ,
     f_{ 03} ,
     f_{ 30} ,
     f_{ 31} ,
     f_{ 32} ,
     f_{ 33}
\right\},
\left\{
    f_{  10} ,
     f_{ 11} ,
     f_{ 12} ,
     f_{ 13} ,
     f_{ 20} ,
     f_{ 21} ,
     f_{ 32} ,
     f_{ 33}
\right\}
\right\}
,
\label{2005-ko-e55-01a}   \\
F_2&=&
\left\{
\left\{
     f_{ 00} ,
     f_{ 10} ,
     f_{ 20} ,
     f_{ 30} ,
     f_{ 03} ,
     f_{ 13} ,
     f_{ 23} ,
     f_{ 33}
\right\},
\left\{
     f_{ 01} ,
     f_{ 11} ,
     f_{ 21} ,
     f_{ 31} ,
     f_{ 02} ,
     f_{ 12} ,
     f_{ 22} ,
     f_{ 32}
\right\}
\right\}
,
\label{2005-ko-e55-01b}   \\
F_3&=&
\left\{
\left\{
     f_{ 01} ,
     f_{ 02} ,
     f_{ 10} ,
     f_{ 13} ,
     f_{ 20} ,
     f_{ 23} ,
     f_{ 31} ,
     f_{ 32}
\right\},
\left\{
     f_{ 00} ,
     f_{ 03} ,
     f_{ 11} ,
     f_{ 12} ,
     f_{ 21} ,
     f_{ 22} ,
     f_{ 30} ,
     f_{ 33}
\right\}
\right\}
,
\label{2005-ko-e55-01c}   \\
F_4&=&
\left\{
\left\{
     f_{ 00} ,
     f_{ 03} ,
     f_{ 11} ,
     f_{ 12} ,
     f_{ 21} ,
     f_{ 22} ,
     f_{ 30} ,
     f_{ 33}
\right\},
\left\{
     f_{ 01} ,
     f_{ 02} ,
     f_{ 10} ,
     f_{ 13} ,
     f_{ 20} ,
     f_{ 23} ,
     f_{ 31} ,
     f_{ 32}
\right\}
\right\}
.
\label{2005-ko-e55-01d}
\end{eqnarray}

Thus any filter which resolves the associated decision problem at once has to
separate
(1)
$\varphi_1$ and $\varphi_3$ from $\varphi_2$ and $\varphi_4$,
(2)
$\varphi_1$ and $\varphi_2$ from $\varphi_3$ and $\varphi_4$,
(3)
$\varphi_2$ and $\varphi_3$ from $\varphi_1$ and $\varphi_4$,
(4)
$\varphi_1$ and $\varphi_4$ from $\varphi_2$ and $\varphi_3$, respectively.

Again, the strategy is to find the unitary transform
\begin{equation}
\textsf{\textbf{U}}'=\frac{1}{2}
\left(
\begin{array}{rrrr}
 1&  1& 1&  1\\
 1& 1& -1& -1\\
1&  -1& 1& -1\\
1& -1& -1&  1
\end{array}
\right) ,
\end{equation}
which yields a basis change
$\textbf{B} \rightarrow \textsf{\textbf{U}}'\textbf{B}=\textbf{B}'$.
Then, measurement of
$\textsf{\textbf{F}}' =
(\textsf{\textbf{U}}')^{-1}\textsf{\textbf{F}}_i\textsf{\textbf{U}}'$
with
\begin{eqnarray}
\textsf{\textbf{F}}_1&=&{\rm diag}(1,0,1,0), \\
\textsf{\textbf{F}}_2&=&{\rm diag}(1,1,0,0),   \\
\textsf{\textbf{F}}_3&=&{\rm diag}(0,1,1,0),   \\
\textsf{\textbf{F}}_4&=&{\rm diag}(1,0,0,1),
\end{eqnarray}
solves the decision problems (D1)--(D4), respectively.
This method can be generalized to more than two arguments in a straightforward manner.

\section{Information spread among quanta}

So why can the parity of a function not be efficiently coded quantum mechanically?
In Ref.~\cite{bbcmw-01}, Beals {\it et al.}
argue that exponential quantum speed-up can be obtained for
partial functions (e.g., problems involving a promise on input~\footnote{
A partial function is a function which is not defined for some of its domain.}),
whereas such speedups cannot be obtained
for any total function.
Another ansatz for an explanation,
put forward by Orus {\it et al.}
in Ref.~\cite{orus-04}, is majorization:
The
probability distribution associated with the quantum state is step-by-step majorized until
it is maximally ordered.
Then a measurement provides the solution with high probability.

We propose here that the lack of efficient quantum algorithms
is due to the nonexistence of
mappings of functions $f$ and decision problems into suitable unitary
transformations
$
\textsf{\textbf{U}}_f
$
which could be used for
a system of states and of filter(s) resolving those states corresponding to that particular
algorithmic problem and no other one.
To give an example, in order for a quantum computation to
resolve the equipartition in Eq.~(\ref{2005-ko-e55-0})
by some equivalent quantum state filter,
any such filter must be based upon an encoding of the functional parity
into some orthogonal set of states.
Thereby, in order for the encoding to be efficient, it should not require the
separate functional evaluation of all classical cases.
On the contrary, the mapping $f \mapsto U_f$, as well as
states and filters need to be conceptualized
in a way which leaves the single functional values {\em undefined},
but concentrates on the structural property of parity alone:
the even or odd number of occurrence of certain functional values (0 or 1)
on the entirety of outputs.
If the filters could resolve singular functional values
in the standard computational basis,
they would essentially model classical information.
Any such state preparation or measurement
would make impossible the encoding of information
``spread among'' multipartite states as mentioned above,
which seems to be one of the advantages of quantum computing.
In this paradigm, entanglement and the suitable superposition of multipartite states
become related concepts, as no multipartite state which can be factored
could be used to ``spread'' information among the quanta (or a group of quanta) corresponding to these factors.

In general, while all classical computable recursive functions
$f$ and decision problems can be coded quantum mechanically, there is no guarantee
that a problem can be coded efficiently by mapping it into the quantum domain.
By an efficient coding of a (binary or $n$-ary) decision
problem  we mean that some quantum circuit
$\textsf{\textbf{U}}_f$ exists
which outputs a state which is uniquely identifiable by a single filter
(or at least by a polynomial number of filters),
the outcome of which corresponds to the solution of this problem.

While the parity of a binary function
of more than one observable has already been mentioned
as an example of quantum computationally ``hard'' problems,
it appears not totally unreasonable to speculate  that functional recursions
and iterations
represent an additional burden on efficiency.
Recursions may require
a space overhead to keep track of the computational path,
in particular if the recursion depth cannot be coded efficiently.
From this point of view,
quantum implementations of the Ackermann or the Busy Beaver functions,
to give just two examples,
may even be less efficient than classical implementations,
where an effective waste management can get rid of many bits;
in particular in the presence of a computable radius of convergence.

\section{State identification and dense coding}
Let us also briefly mention another issue related to state identification
if there is a mismatch between the context in which
information is prepared and a different context, in which this information
is retrieved.
Based on such a context mismatch,
a ``dense coding'' scheme has been proposed \cite{581773}
to probabilistically encode ``more'' than one
classical bits into one quantum bit (despite Holevo's bound).
This method is based on the fact that the qubit states
$\vert 0\rangle$
and
$\vert 1\rangle$
span the computational basis
$\{ (1,0)^T,(0,1)^T\}$, as already mentioned before,
and that any coding of a qubit state which is neither orthogonal nor
collinear, such as $(\cos (\pi /8),\sin (\pi /8))^T$,
results in a probability of detecting it in the
original states governed by its projection onto them.
The argument is about efficiency of state identification in the classical
and quantum case for ``misaligned'' systems of states.

Alas, when speaking about coding and representation efficiency of statistical raw data,
it is mandatory to take an issue into account which changes the classical
framework rather dramatically.
As has been pointed out repeatedly by Summhammer
\cite{sum-1,summi:93},  the ``true'' probability of the
occurrence of a (classical) bit is unknown.
Frequency counts are just approximations to this value.
As it turns out, if a {\em finite} amount of information is used
to characterize the probability $p$ by the actually observed relative frequencies
$L/N$, where $N$ is the number of experiments and $L$ is the number of occurrences,
then the accuracy varies as a function of $p$.
Thus,
a representation of the data
has to be chosen which guarantees a constant rate of accuracy over
the entire probability range.
This results in a redefinition of the functional representation of the
relative frequency which is very similar to the quantum mechanical
representation by vectors and projectors in Hilbert space.
(Compare Mermin's
representation~\cite{mermin-02,mermin-04,mermin-qc}
of classical information theory and reversible operations on classical bits
in linear vector spaces in some analogy to the quantum formalism.)
From this point of view, taking the finite coding of probabilities by
relative frequencies into account, the classical and the quantum ``dense''
coding schemes become equivalent.

\section{Summary}

We have presented an analysis of quantum computations in terms of state identification
whose complexity grows linearly with the number of bits.
Thereby, we have characterized this domain by partitions of state space,
as well as by unitary transformations of the associated filter systems.
Such systems are not bound by the individual classical values,
as information about the (parallelized)
result of a computation may be ``spread among'' the quanta
in a way which makes it impossible to reconstruct the result
by measuring the quanta separately.
At the same time, such distributed information could be analyzed a single (or a few)
measurement(s) by proper filters resolving the computed proposition.

The method does not yield a constructive, operational method
for deciding
whether or not (and if so, how) functions or decision problems of practical interest
can be efficiently coded into quantum algorithms.
From a foundational point of view it is interesting
to realize that, while every suitable equipartitioning of state space
is equivalent to some proposition which can be interpreted as an outcome of some
quantum computation,
not all decision problems or functional evaluations which can be rephrased as
state partitions can be translated efficiently into the quantum domain.

\section*{Acknowledgments}
I am grateful to David Mermin for pointing out a generalization
of a two-bit problem to functional parity.


\begin{thebibliography}{10}
\bibitem{Ozhigov:1997}
Ozhigov, Y., Quantum computer can not speed up iterated applications of a black
  box.
\end{thebibliography}

\end{document}